\documentclass[showpacs,eqsecnum,prd,preprint]{revtex4-1}
\usepackage[dvips]{graphicx}
\usepackage{amsmath,amssymb}

\begin{document}

\begin{flushright}
EPHOU-16-012
\end{flushright}

\title{Finite-Size Corrections to the Excitation Energy Transfer in a Massless Scalar Interaction Model}

\author{Nobuki Maeda,$^{1}$ Tetsuo Yabuki,$^2$ Yutaka Tobita,$^3$ and Kenzo Ishikawa$^1$}
\affiliation{$^1$Department of Physics, Hokkaido University, Sapporo 060-0810, Japan\\
$^2$Department of Environmental and Symbiotic Sciences, Rakuno Gakuen University, Ebetsu 069-8501, Japan\\
$^3$Department of Mathematics and Physics, Hirosaki University, Hirosaki 036-8561, Japan
}

\date{\today}

\begin{abstract}
We study the excitation energy transfer (EET) for a simple model in which a massless scalar particle is
exchanged between two molecules. 
We show that a finite-size effect appears in EET by the interaction energy due to overlapping of the quantum waves 
in a short time interval. 
The effect generates finite-size corrections to Fermi's golden rule and modifies EET probability from the standard formula in the F\"orster mechanism. 
The correction terms come from transition modes outside the resonance energy region and enhance EET probability substantially. 
\end{abstract}
%\pacs{w-07}

\maketitle

\section{Introduction}

Excitation energy transfer (EET) phenomena between molecules by exchange of a virtual photon 
play an important role in science and technology, such as photosynthesis and bio-sensors. 
EET occurs as a molecule (donor) makes the transition from an excited state to a lower-energy state 
and a neighboring molecule (acceptor) is excited to a higher-energy state by exchange of a photon. 
The transfer time between the two molecules that are a few nanometers away is less than a few picoseconds. 

Recent development of experimental technology (x-ray crystallography, ultrafast spectroscopy, etc.) 
makes it possible to clarify considerably the microscopic mechanism of EET in the photosynthesis system. 
In conventional theory, the static Coulomb potential between the two molecules is used with Fermi's golden rule \cite{dirac} 
for the calculation of EET \cite{oppen,oppen2,fret,dex}, and global features of EET were understood.  
The F\"orster mechanism \cite{fret} is the standard theory for EET in the weak-coupling regime 
in which the back transfer from the acceptor to the donor is suppressed by quantum decoherence due to the interaction with the environment. 
In this mechanism, the dipole--dipole potential, which comes from the direct Coulomb term, is used in the point-particle limit.  
However, this standard theory fails to explain the extremely high efficiency of EET in the photosynthesis system, 
such as the light-harvesting complex of purple bacteria. 
Many approaches to this problem have been proposed in experimental and theoretical works before now, 
and have revealed the important role of the exciton modes in the enhancement of the EET rate \cite{p0,p1,p2,p3,p4,p5,p6,p7,ishizaki,schulten}. 
These works focused on the EET in the resonance energy region. 

We investigate a new possibility for solving the high efficiency problem by taking into account the effect of the photon 
exchange between the molecules. 
The interaction energy due to overlapping between the photon and the electron waves causes a finite-size correction to 
Fermi's golden rule \cite{ishikawa,ishikawa1,ishikawa2}. 
This correction has been mostly overlooked until recently. 
We study a simple model in which EET is caused by exchange of a massless scalar particle, 
which is an analog of the photon, and  
electron wave functions are assumed to be one-particle states and a Gaussian function. 
In the transition probability, in addition to the resonance energy peak due to Fermi's golden rule, 
a broad bump due to the finite-size correction appears outside the resonance energy region. 
The resonance peak is hardly dependent on the molecular size except for the dipole moment dependence, but  
the broad bump is strongly dependent on the molecular size. 
While the resonance peak always increases with the time interval, 
the broad bump rapidly increases for a short time interval and remains constant for a long time interval. 

The extremes of the rapid-rise and constant behavior are characteristics of the finite-size effect. 
The finite-size correction enhances not the transition {\it rate} but the transition {\it probability} by the constant correction term 
for a long time interval. 
Thus, the energy transfer through the correction term is totally different from the standard one in the resonance energy region. 
The estimation in a photosynthesis system indicates that the transition probability for EET could be enhanced by the finite-size effect. 

The continuum modes of the scalar-particle cause the transition to the broad bump energy region. 
The broad bump and the rapid-rise behavior are related to the energy-time complementarity. 
We use time-dependent perturbation theory, and the finite-size effect appears at the tree level. 
Since the rapid-rise time interval is very short, the back transfer can be ignored in the perturbation theory,  
which is usually used in the weak-coupling regime. 
We verify the validity of the perturbation theory by calculating the higher-order effect. 
We estimate the lifetime effect and the renormalization of the finite-size correction in the higher-order perturbation.  
It is shown that the natural line width is hardly dependent on the finite-size correction. 

The paper is organized as follows. 
We review the basic concept of the finite-size corrections to Fermi's golden rule in Sec.~{\ref{basic}}. 
A massless scalar interaction model is introduced, and the spontaneous emission probability is obtained in Sec.~{\ref{massless}}. 
The transition amplitude in EET is derived in the second-order perturbation theory in Sec.~\ref{transitionam}. 
We discuss the relation between our theory and standard theory in Sec.~{\ref{relation}}. 
Various numerical results for EET are given in Sec.~\ref{numerical}. 
A summary is given in Sec.~\ref{summary}. 
In this paper, we use the natural unit ($\hbar=c=1$). 

\section{Basic concept for finite-size corrections to Fermi's golden rule}\label{basic}

In this section, we review a basic concept in the quantum mechanics underlying our theory. 
We consider a Hamiltonian that is decomposed into a free Hamiltonian $H_0$ and 
an interaction Hamiltonian $H_{\rm int}$ as
\begin{equation}
H=H_0+H_{\rm int}. 
\end{equation}
We assume that $H_0$ does not change the number of particles but that $H_{\rm int}$ does. 
For example, $H_{\rm int}$ creates or annihilates a massless particle. 
Let us introduce energy eigenstates of the free Hamiltonian as
\begin{equation}
H_0\vert n\rangle=E_n\vert n\rangle,\ \langle n\vert n'\rangle=\delta_{nn'}.
\end{equation}
We assume that the eigenstates of $H_0$ are also particle number eigenstates. 
Then, we have the relation
\begin{equation}
\langle n\vert H\vert n'\rangle=E_n\delta_{nn'}+\langle n\vert H_{\rm int}\vert n'\rangle(1-\delta_{nn'}). 
\label{off}
\end{equation}
Time evolution of the quantum state is governed by Schr\"odinger's wave equation,
\begin{equation}
i\frac{d}{dt}\vert \psi(t)\rangle=H\vert \psi(t)\rangle.
\label{schr}
\end{equation}
We consider a nonstationary wave function that is expanded by energy eigenstates as
\begin{equation}
\vert \psi(t)\rangle=c_0(t)e^{-iE_0 t}\vert 0\rangle+\sum_{n\neq0}c_n(t)e^{-iE_n t}\vert n\rangle,\ c_n(0)=\delta_{n0},
\label{nonstasta}
\end{equation}
where $\vert 0\rangle$ is the initial state and $c_n(t)$ is the transition amplitude for the final state $\vert n\rangle$. 
Following von Neumann's fundamental principle of quantum mechanics \cite{von}, the transition probability 
to the state $\vert n\rangle$ at $t=T$ is given by $P_n(T)=\vert\langle n\vert\psi(T)\rangle\vert^2=\vert c_n(T)\vert^2$. 
The transition amplitude $c_n(T)$ is given by
\begin{equation}
c_n(T)=\langle n\vert \mathcal{T}e^{-i\int_0^T H_{\rm int}^I(t)dt}\vert0\rangle,
\label{timeorder}
\end{equation}
where $H^I_{\rm int}(t)=e^{iH_0 t}H_{\rm int}e^{-iH_0 t}$ and 
$\mathcal{T}$ stands for the time-ordering product. 
In the time-dependent perturbation approximation, the transition amplitude is calculated 
by expanding the exponential in the above equation. 

The energy conservation law is written as
\begin{equation}
\frac{d}{dt}\langle\psi(t)\vert H\vert \psi(t)\rangle=0,
\end{equation}
which is easily proven by using Eq.~(\ref{schr}). 
The orthogonality condition is also conserved as $\frac{d}{dt}\langle n(t)\vert n'(t)\rangle=0$ for $\vert n(t)\rangle=e^{-iHt}\vert n\rangle$. 
By using the normalization condition $\langle\psi(t)\vert\psi(t)\rangle=1$ and Eq.~(\ref{off}), the expectation value of the total energy at 
$t=T$ is written as
\begin{equation}
\langle\psi(T)\vert H\vert\psi(T)\rangle=E_0+\sum_{n\neq0}(E_n-E_0)\vert c_n(T)\vert^2 + \sum_{n\neq n'}c^*_n(T) c_{n'}(T)e^{i(E_n-E_{n'})T}\langle n\vert H_{\rm int}
\vert n'\rangle.
\label{expect}
\end{equation}
Since the expectation value of the total energy at $t=0$ is 
$\langle\psi(0)\vert H\vert\psi(0)\rangle=E_0$, the energy conservation law means that the second term and third term cancel each other out 
in the right-hand side of Eq.~(\ref{expect}). 
The second term is the expectation value of the energy increase for $H_0$ at $t=T$ and 
corresponds to finite-size corrections to Fermi's golden rule. 
The third term is the expectation value of $H_{\rm int}$, which is the overlap of different final states through $H_{\rm int}$ 
and corresponds to the interference term due to diffraction. 
Since $H_{\rm int}$ changes a particle number, the overlap occurs at the region where a particle is emitted or absorbed.  
In these regions, the wave nature of quanta remains and the interaction cannot be disregarded.  

For large $T$, the energy for $H_0$ is conserved approximately and the transition amplitude is dominant for  
$\vert E_n-E_0\vert\lesssim2\pi/T$, which we call the resonance energy region. 
In this region, the transition probability $P$ is approximately proportional to $T$ as $P=\Gamma T$. 
Then, the transition rate can be calculated by Fermi's golden rule in the resonance energy region. 
However, a finite-size correction $P^{(d)}$, which hardly depends on $T$, could appear 
outside the resonance energy region as $P=\Gamma T+P^{(d)}$ for a large but finite $T$ \cite{ishikawa,ishikawa1,ishikawa2}. 
The transition probability rapidly increases to $P^{(d)}$ at small $T$. 
This rapid rise time-region corresponds to the overlapping time of the wave functions. 
This rapid time dependence implies a broad energy spectrum contributing to $P^{(d)}$. 
In the present paper, we investigate the property of the finite-size correction for EET phenomena in a simple model. 

\section{Massless scalar interaction model}
\label{massless}
We consider electron states in molecules as bound states in confining potentials induced by nuclei. 
We assume that the interaction between the bound states is generated by a massless scalar field, 
which is an analog of the photon field. 
The model Hamiltonian is given by
\begin{eqnarray}
H&=&H_0+H_{\rm int},\\
H_0&=&H^A_0+H^D_0+H_0^\phi,\nonumber\\
H^A_0&=&\frac{p_A^2}{2m}+V_A(\boldsymbol{r}_A,q_A)+H^A_{\rm nucl}(q_A),\nonumber\\
H^D_0&=&\frac{p_D^2}{2m}+V_D(\boldsymbol{r}_D,q_D)+H^D_{\rm nucl}(q_D),\nonumber\\ 
H_0^\phi&=&\int\frac{d^3 k}{(2\pi)^3}E_k a^\dagger_k a_k,\ 
H_{\rm int}=g\phi(\boldsymbol{r}_A)+g\phi(\boldsymbol{r}_D),\nonumber
\end{eqnarray}
where $\boldsymbol{r}_A$, $\boldsymbol{p}_A$ and $\boldsymbol{r}_D$, $\boldsymbol{p}_D$ are electron coordinates and momenta, 
and $V_A$ and $V_D$ are confining potentials for electrons in molecules $A$ (Acceptor) and $D$ (Donor) respectively, 
$E_k=k$, $[a_k,a^\dagger_{k'}]=(2\pi)^3\delta^3(k-k')$, $m$ is the electron mass, and $g$ is a coupling constant. 
We assume that electron states are one-particle states that are well separated and regarded as two distinct particles. 
THe terms $H^A_{\rm nucl}(q_A)$ and $H^D_{\rm nucl}(q_D)$ are the Hamiltonian for the nuclear coordinates $q_A$ and $q_D$. 
Details of confining potentials and nuclear Hamiltonians are not necessary for the present calculation. 
The scalar field $\phi$ is expanded as
\begin{equation}
\phi(\boldsymbol{r})=\int\frac{d^3 k}{(2\pi)^3}\frac{1}{\sqrt{2k}}(a_k e^{i\boldsymbol{k\cdot r}}+a^\dagger_k e^{-i\boldsymbol{k\cdot r}}).
\end{equation}

We assume that the nuclear mass is heavy enough to treat the nuclear coordinates $q_A$ and $q_D$ as 
the adiabatic coordinates in the Born-Oppenheimer approximation. 
Then, the initial state $\vert \psi_i \rangle$ and the final state  $\vert \psi_f \rangle$ for EET are given by 
\begin{eqnarray}
\vert \psi_i\rangle&=&\vert 0\rangle\vert A0\rangle\vert{\rm nucl}\rangle_{A0}\vert D1\rangle\vert{\rm nucl}\rangle_{D1},\label{initialfinal}\\
\vert \psi_f\rangle&=&\vert 0\rangle\vert A1\rangle\vert{\rm nucl}\rangle_{A1}\vert D0\rangle\vert{\rm nucl}\rangle_{D0}\nonumber
\end{eqnarray}
where $\vert 0\rangle$ is the number 0 state for $\phi$; 
$\vert A0\rangle$,  $\vert D0\rangle$ and $\vert A1\rangle$,  $\vert D1\rangle$ are ground states and excited states 
for electrons in molecules $A$ and $D$ respectively; and 
$\vert{\rm nucl}\rangle_{A0}$, $\vert{\rm nucl}\rangle_{D0}$ and $\vert{\rm nucl}\rangle_{A1}$, $\vert{\rm nucl}\rangle_{D1}$ are 
nuclear states for ground states and excited states of electrons in molecules $A$ and $D$ respectively. 
We ignore the nuclear coordinate dependence of the electron states (Condon approximation). 
The electron states satisfy orthogonality: $\langle A1\vert A0\rangle=\langle D0\vert D1\rangle=0$. 
The overlap of nuclear states $S_A=_{A1}\!\!\langle{\rm nucl}\vert{\rm nucl}\rangle_{A0}$ and 
$S_D=_{D0}\!\!\langle{\rm nucl}\vert{\rm nucl}\rangle_{D1}$ are called Franck-Condon factors. 
Then, $\vert \psi_i\rangle$ and $\vert \psi_f\rangle$ are regarded as eigenstates of $H_0$ approximately. 

Let us introduce the energy eigenvalues of $H_0^A$, $H_0^D$ as $E_A^0$, $E_D^0$ for ground states and 
$E_A^1$, $E_D^1$ for excited states in the molecules $A$ and $D$ respectively. 
That is,
\begin{eqnarray}
H_0^A\vert A0\rangle\vert{\rm nucl}\rangle_{A0}&=&E_A^0\vert A0\rangle\vert{\rm nucl}\rangle_{A0},\label{eigen}\\
H_0^A\vert A1\rangle\vert{\rm nucl}\rangle_{A1}&=&E_A^1\vert A1\rangle\vert{\rm nucl}\rangle_{A1},\nonumber\\ 
H_0^D\vert D0\rangle\vert{\rm nucl}\rangle_{D0}&=&E_D^0\vert D0\rangle\vert{\rm nucl}\rangle_{D0},\nonumber\\
H_0^D\vert D1\rangle\vert{\rm nucl}\rangle_{D1}&=&E_D^1\vert D1\rangle\vert{\rm nucl}\rangle_{D1}.\nonumber
\end{eqnarray} 
The energy eigenvalues of $\vert \psi_i\rangle$ and $\vert \psi_f\rangle$ are given by $E_A^0+E_D^1$ and 
$E_A^1+E_D^0$ respectively. 
Transition energies are defined as $E_A=E_A^1-E_A^0$ and $E_D=E_D^1-E_D^0$. 
Then, the energy change from $\vert \psi_i\rangle$ to $\vert \psi_f\rangle$ is given by $E_A-E_D$. 
We assume that corresponding energy eigenstates for electrons in the molecules are given by
\begin{eqnarray}
\langle\boldsymbol{r}_A\vert A0\rangle&=&N_{A0}e^{-\sigma_A^2 (\boldsymbol{r}_A-\bar{\boldsymbol{r}}_A)^2/2},
\label{wf}\\
\langle\boldsymbol{r}_A\vert A1\rangle&=&N_{A1}\hat{\boldsymbol{\mu}}_A\cdot (\boldsymbol{r}_A-\bar{\boldsymbol{r}}_{A})
e^{-\sigma_A^2 (\boldsymbol{r}_A-\bar{\boldsymbol{r}}_A)^2/2}, \nonumber\\
\langle\boldsymbol{r}_D\vert D0\rangle&=&N_{D0}e^{-\sigma_D^2 (\boldsymbol{r}_D-\bar{\boldsymbol{r}}_D)^2/2},\nonumber\\
\langle\boldsymbol{r}_D\vert D1\rangle&=&N_{D1}\hat{\boldsymbol{\mu}}_D\cdot (\boldsymbol{r}_D-\bar{\boldsymbol{r}}_{D})
e^{-\sigma_D^2 (\boldsymbol{r}_D-\bar{\boldsymbol{r}}_D)^2/2},\nonumber
\end{eqnarray} 
where $\hat{\boldsymbol{\mu}}_A$ and $\hat{\boldsymbol{\mu}}_D$ are unit vectors for the transition dipole moments, and  
$\bar{\boldsymbol{r}}_A$, $\bar{\boldsymbol{r}}_D$ and $1/\sigma_A$, $1/\sigma_D$ correspond to positions and sizes 
of the molecules $A$ and $D$ respectively. 
Normalization constants are given by
\begin{eqnarray}
N_{A0}&=&(\sigma_A/\sqrt{\pi})^{3/2},\ N_{A1}=\sqrt{2}\sigma_A(\sigma_A/\sqrt{\pi})^{3/2},\\
N_{D0}&=&(\sigma_D/\sqrt{\pi})^{3/2},\ N_{D1}=\sqrt{2}\sigma_D(\sigma_D/\sqrt{\pi})^{3/2}.\nonumber
\end{eqnarray} 
The transition dipole moments are calculated as
\begin{eqnarray}
\mu^i_A&=&g\int d^3r_A\langle A1\vert\boldsymbol{r}_A\rangle r_A^i\langle\boldsymbol{r}_A\vert A0\rangle=
\hat{\mu}_A^ig/\sqrt{2}\sigma_A,\\
\mu^i_D&=&g\int d^3r_D\langle D0\vert\boldsymbol{r}_D\rangle r_D^i\langle\boldsymbol{r}_D\vert D1\rangle=
\hat{\mu}_D^ig/\sqrt{2}\sigma_D.\nonumber
\end{eqnarray} 
The dipole approximation corresponds to taking the limit $\sigma_A$, $\sigma_D\rightarrow\infty$ with $\mu_A$ and $\mu_D$ fixed. 
Note that the transition dipole moments are $O(g)$. 

\subsection{Nonstationary state}
\label{nonstat}
In this section, we study the property of the nonstationary state of the present model according to 
Sec.~\ref{basic}. 
The nuclear states are not considered here for simplicity. 
We obtain a nonstationary state in Eq.~(\ref{nonstasta}) up to $O(g^2)$ in the present model as
\begin{equation}
\vert \psi(t)\rangle=(1+c_i^{(2)}(t))e^{-i(E_A^0+E_D^1) t}\vert \psi_i\rangle+\int\frac{d^3k}{(2\pi)^3}
c^{(1)}_{\boldsymbol k}(t)e^{-i(k+E_A^0+E_D^0) t}
\vert\psi_{\boldsymbol k}\rangle
+c_f^{(2)}(t)e^{-i(E_A^1+E_D^0) t}\vert \psi_f\rangle
\end{equation}
where $\vert \psi_i\rangle=\vert 0\rangle\vert A0\rangle\vert D1\rangle$, 
$\vert\psi_{\boldsymbol k}\rangle=\vert \boldsymbol k\rangle\vert A0\rangle\vert D0\rangle$, and 
$\vert \psi_f\rangle=\vert 0\rangle\vert A1\rangle\vert D0\rangle$. 
Also, $\vert 0\rangle$ is the number 0 state and $\vert\boldsymbol{k}\rangle$ is a one-particle state with 
a wave number $\boldsymbol{k}$ for the scalar particle $\phi$. 
We consider only main states in the present paper and omit the other states, i.e., two-particle emission state, 
$\vert {\boldsymbol k}\rangle\vert A1\rangle\vert D1\rangle$, etc.  
The time-dependent coefficient $c_i^{(2)}(t)$ is $O(g^2)$ and
$c^{(1)}_{\boldsymbol k}(t)$ is a spontaneous particle emission amplitude of $O(g)$. 
The term $\vert \psi_f\rangle$ is the final state for EET and $c_f^{(2)}(t)$ is the EET transition amplitude of $O(g^2)$. 

Since $\vert \psi(0)\rangle=\vert \psi_i\rangle$ at $t=0$, 
the initial condition is given by $c_i^{(2)}(0)=c^{(1)}_{\boldsymbol k}(0)=c_f^{(2)}(0)=0$. 
Time evolution of the nonstationary state is governed by Schr\"odinger's wave equation (\ref{schr}). 
The transition amplitude for the spontaneous scalar-particle emission, $c^{(1)}_{\boldsymbol k}(T)$, 
is given by
\begin{eqnarray}
c^{(1)}_{\boldsymbol{k}}(T)&=&-i\int_0^T dt \langle \psi_{\boldsymbol k}\vert H_{\rm int}^I(t)\vert \psi_i\rangle
=-i\int_0^T dt e^{i(k-E_D)t}\langle \psi_{\boldsymbol k}\vert H_{\rm int}\vert \psi_i\rangle\nonumber\\
&=&-ie^{i\frac{k-E_D}{2}T}\frac{2\sin(\frac{k-E_D}{2}T)}{k-E_D}\langle \psi_{\boldsymbol k}
\vert H_{\rm int}\vert \psi_i\rangle
\label{oneamp}
\end{eqnarray}
where $H_{\rm int}^I(t)=e^{iH_0 t}H_{\rm int}e^{-iH_0 t}$. 

We obtain the relation
\begin{equation}
\vert 1+c_i^{(2)}(t)\vert^2+\int\frac{d^3k}{(2\pi)^3}\vert c^{(1)}_{\boldsymbol k}(t)\vert^2=1,
\label{norma}
\end{equation}
up to $O(g^2)$. 
This relation corresponds to the normalization condition $\langle\psi(t)\vert\psi(t)\rangle=1$. 

The expectation value for $H_0$ at $t=T$ is given by
\begin{eqnarray}
\langle\psi(T)\vert H_0\vert\psi(T)\rangle&=&(E_A^0+E_D^1) \vert 1+c_i^{(2)}(T)\vert^2+
\int\frac{d^3k}{(2\pi)^3}(k+E_A^0+E_D^0)\vert c^{(1)}_{\boldsymbol k}(T)\vert^2\nonumber\\
&=&E_A^0+E_D^1+\int\frac{d^3k}{(2\pi)^3}(k-E_D)\vert c^{(1)}_{\boldsymbol k}(T)\vert^2
\end{eqnarray}
up to $O(g^2)$, where we used Eq.~(\ref{norma}). 
The initial energy is $E_A^0+E_D^1$. 
Therefore, the energy increase for $H_0$ at $t=T$ is given by
\begin{equation}
\int\frac{d^3k}{(2\pi)^3}(k-E_D)\vert c^{(1)}_{\boldsymbol k}(T)\vert^2
=\int\frac{d^3k}{(2\pi)^3}(k-E_D)
\left(\frac{2\sin(\frac{k-E_D}{2}T)}{k-E_D}\right)^2
\vert\langle \psi_{\boldsymbol k}
\vert H_{\rm int}\vert \psi_i\rangle\vert^2,
\label{increase}
\end{equation}
where we used Eq.~(\ref{oneamp}). 
In the following section, we show that the transition probability for $k>E_D$ contributes to 
the energy increase of $H_0$ for a finite $T$. 

The expectation value for $H_{\rm int}$ at $t=T$ is given by
\begin{eqnarray}
\langle\psi(T)\vert H_{\rm int}\vert\psi(T)\rangle&=&
\int\frac{d^3k}{(2\pi)^3}\{c^{(1)}_{\boldsymbol k}(T)e^{-i(k-E_D) T}
\langle \psi_i\vert H_{\rm int}\vert\psi_{\boldsymbol k}\rangle+c^{(1)*}_{\boldsymbol k}(t)e^{i(k-E_D) T}
\langle \psi_{\boldsymbol k}\vert H_{\rm int}\vert\psi_i\rangle\}\nonumber\\
&=&-\int\frac{d^3k}{(2\pi)^3}
\frac{\{2\sin(\frac{k-E_D}{2}T)\}^2}{k-E_D}
\vert\langle \psi_{\boldsymbol k}
\vert H_{\rm int}\vert \psi_i\rangle\vert^2,
\end{eqnarray}
up to $O(g^2)$. 
Here, we used Eq.~(\ref{oneamp}). 
The above interaction energy between the initial state and the emission state 
cancels out the energy increase for $H_0$ in Eq.~(\ref{increase}). 
Therefore, it is shown that the total energy $H_0+H_{\rm int}$ is conserved up to $O(g^2)$. 
Using a similar procedure, we can prove energy conservation in a higher order of $g$. 

\subsection{Finite-size correction of the spontaneous scalar-particle emission}
\label{spon}
Let us calculate the probability of spontaneous scalar-particle emission for the transition of $\vert \psi_i\rangle
\rightarrow\vert\psi_{\boldsymbol k}\rangle$. 
The nuclear states are not considered here for simplicity. 
Using Eqs.~(\ref{wf}) and (\ref{oneamp}), we obtain the transition amplitude as
\begin{eqnarray}
c^{(1)}_{\boldsymbol{k}}(T)&=&
-ie^{i\frac{k-E_D}{2}T}\frac{2\sin(\frac{k-E_D}{2}T)}{k-E_D}\langle \psi_{\boldsymbol k}
\vert g\phi(\boldsymbol{r}_D)\vert \psi_i\rangle
\\
&=&-e^{i\frac{k-E_D}{2}T}\frac{2\sin(\frac{k-E_D}{2}T)}{k-E_D}\sqrt{\frac{k}{2}}(\boldsymbol{\mu}_D\cdot\frac{\boldsymbol{k}}{k})
e^{-\frac{k^2}{4\sigma_D^2}-i\boldsymbol{k}\cdot
\boldsymbol{\bar r}_D},
\nonumber
\end{eqnarray}
and the transition probability is given by
\begin{equation}
\vert c^{(1)}_{\boldsymbol{k}}(T)\vert^2=\frac{k}{2}(\boldsymbol{\mu}_D\cdot\frac{\boldsymbol{k}}{k})^2e^{-\frac{k^2}{2\sigma_D^2}}
\left(\frac{2\sin(\frac{k-E_D}{2}T)}{k-E_D}\right)^2.
\end{equation}
The transition probability without observing the emitted scalar particle is given by summing $\boldsymbol{k}$ as
\begin{equation}
P_{\rm rad}(T)=\int \frac{d^3k}{(2\pi)^3}\vert c^{(1)}_{\boldsymbol{k}}(T)\vert^2
=\frac{\mu_D^2}{6\pi}E_D^2\int_0^\infty \frac{dk}{2\pi E_D}f_{\rm rad}(k,T),
\end{equation}
where 
\begin{equation}
f_{\rm rad}(k,T)=\frac{k^3}{E_D} e^{-\frac{k^2}{2\sigma_D^2}}\left(\frac{2\sin(\frac{k-E_D}{2}T)}{k-E_D}\right)^2.
\end{equation}
This equation shows that $\sigma_D$ behaves as a cutoff for the higher energy of the emitted scalar particle. 

\begin{figure}[htb]
\begin{center}
\includegraphics[width=9cm]{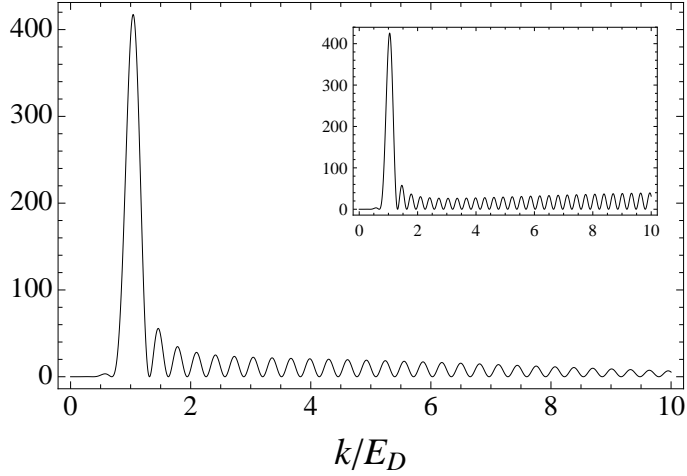}
\caption{The figure shows $f_{\rm rad}(k,T)$ for $T E_D=20$, $\sigma_D/E_D=5$ (=15 for the inset), and $0<k/E_D<10$. }
\label{fig_emission}
\end{center}
\end{figure}

In Fig.~\ref{fig_emission}, $f_{\rm rad}(k,T)$ is plotted. 
The peak at $k/E_D=1$ is the resonance peak and the broad bump at $k/E_D>1$ is the finite-size correction. 
This broad bump increases the expectation value of the energy for $H_0$. 
The height of the peak is proportional to $T^2$ and the broad bump is almost independent of $T$ except for 
the rapid oscillation. 
As shown in the inset of Fig.~\ref{fig_emission}, the bump becomes wide for large $\sigma_D$. 
The width of the bump is roughly estimated by $\Delta E\approx\sigma_D$. 
In the limit of $T\rightarrow\infty$, the spontaneous emission rate becomes
\begin{equation}
\Gamma_{\rm rad}=\lim_{T\rightarrow\infty}\frac{P_{\rm rad}(T)}{T}=\frac{\mu_D^2}{6\pi}E_D^3 e^{-\frac{E_D^2}{2\sigma_D^2}}.
\label{gammarad}
\end{equation}
Then, in the limit of $\sigma_D\rightarrow\infty$ (dipole approximation), the spontaneous emission rate becomes 
$(\mu_D^2/6\pi)E_D^3$, which corresponds to Einstein's A coefficient for spontaneous photon emission.

\begin{figure}[htb]
\begin{center}
\includegraphics[width=9cm]{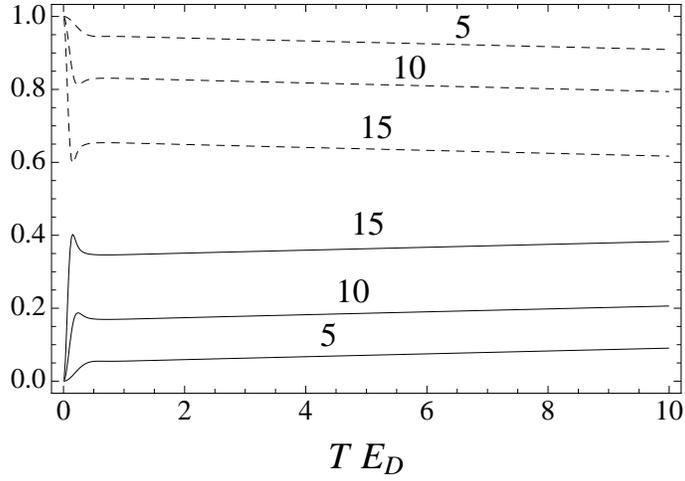}
\caption{The figure shows $P_{\rm rad}(T)$ (solid lines) and $P_i(T)=1-P_{\rm rad}(T)$ (dashed lines) are plotted for 
$\sigma_D/E_D=5$, 10, 15, $E_D/\Gamma_{\rm rad}=250$, and $0<T E_D<10$. }
\label{fig_emission2}
\end{center}
\end{figure}

In Fig.~\ref{fig_emission2}, the transition probability $P_{\rm rad}(T)$ and the survival probability $P_i(T)=1-P_{\rm rad}(T)$ are plotted for  $E_D/\Gamma_{\rm rad}=250$. 
As seen in Fig.~\ref{fig_emission2}, $P_{\rm rad}(T)$ increases to $P^{(d)}_{\rm rad}$ rapidly at small $T$ and is 
approximately given by
\begin{equation}
P_{\rm rad}(T)=\Gamma_{\rm rad}T+P^{(d)}_{\rm rad}
\end{equation}
for $T E_D>1$. 
The rapid-rise time region of the probability is roughly estimated by $\Delta t\approx 1/\Delta E\approx 1/\sigma_D$, where $\Delta E$ is the width of the bump. 
Note that $\Delta t$ corresponds to the overlapping time of electrons and scalar-particle wave functions. 
For $\sigma_D/E_D=15$, the typical time is estimated as $P_{\rm rad}^{(d)}=86.3\Gamma_{\rm rad}/E_D=0.35$, where 
$P_{\rm rad}^{(d)}$ is a finite-size correction to Fermi's golden rule, $P_{\rm rad}(T)=\Gamma_{\rm rad}T$. 
If we scale the parameter as $E_D/\Gamma_{\rm rad}=250 \alpha$, then the correction becomes $P_{\rm rad}^{(d)}=0.35/\alpha$.  
For a large scale parameter $\alpha$, $P_{\rm rad}^{(d)}\ll1$ and the lowest-order perturbation is good. 
For a small scale parameter $\alpha$, $P_{\rm rad}^{(d)}$ becomes larger than 1.  
In this case, the lowest-order perturbation is not good and the higher-order effect must be included. 
In the next section, we show that the finite-size correction $P_{\rm rad}^{(d)}$ is 
renormalized as $\tilde{P}_{\rm rad}^{(d)}=P_{\rm rad}^{(d)}/(1+P_{\rm rad}^{(d)})$ in the higher-order approximation \cite{ww,ww2,ishikawa3,ishikawa4}. 

The result obtained in this section means that the emitted scalar particle has a higher energy than 
the donor excitation energy $E_D$ due to the interaction energy. 
This suggests that the acceptor energy delivered by the scalar particle in EET phenomena 
has a higher energy than $E_D$ due to the interaction energy. 
Actually, in the following sections, it is shown that similar bump structure appears at a higher energy than $E_D$ in 
EET calculation.

\subsection{Higher-order corrections}
\label{lifetime}
In this section, we estimate the finite-size effect and the lifetime effect in the higher-order corrections. 
We use Weisskopf--Wigner theory \cite{ww,ww2} to include the higher-order corrections. 
For simplicity, we consider only a Hilbert space spanned by
$\vert \psi_i\rangle=\vert 0\rangle\vert D1\rangle$ and 
$\vert\psi_{\boldsymbol k}\rangle=\vert \boldsymbol k\rangle\vert D0\rangle$. 
The interaction Hamiltonian $H_{\rm int}$ is assumed to be projected into this restricted Hilbert space. 
The nonstationary state is given by
\begin{equation}
\vert \psi(t)\rangle=c_i(t)e^{-iE_D^1 t}\vert \psi_i\rangle+\int\frac{d^3k}{(2\pi)^3}
c_{\boldsymbol k}(t)e^{-i(k+E_D^0) t}\vert\psi_{\boldsymbol k}\rangle.
\end{equation}
The initial conditions are $c_i(0)=1$ and $c_{\boldsymbol k}(0)=0$. 
Using Eq.~(\ref{timeorder}), we obtain the following differential equations for the coefficients. 
\begin{eqnarray}
\frac{dc_i(t)}{dt}&=&-i\int\frac{d^3k}{(2\pi)^3}
\langle \psi_i
\vert H_{\rm int}\vert \psi_{\boldsymbol k}\rangle 
e^{-i(k-E_D)t}c_{\boldsymbol k}(t),\label{eqa}\\
\frac{dc_{\boldsymbol k}(t)}{dt}&=&-i\langle \psi_{\boldsymbol k}
\vert H_{\rm int}\vert \psi_i\rangle 
e^{-i(E_D-k)t} c_i(t).\label{eqb}
\end{eqnarray}

In order to include the finite-size effect and the lifetime effect, we use the following ansatz \cite{ishikawa3,ishikawa4}, 
which is derived by considering the higher-order perturbation theory: 
\begin{equation}
c_i(t)=A(t)e^{-\gamma t/2},
\label{ansatz}
\end{equation}
where $A(t)$ ($\leq1$) satisfies $A(0)=1$ and $A(t)=A^{(d)}$ for $t\geq\Delta t$. 
Here, $A^{(d)}$ and $\gamma$ are constants and $\Delta t$ is the rapid-rise time interval ($\Delta t\ll 1/E_D$). 
Substituting this ansatz into Eq.~(\ref{eqb}) and integrating both sides of the equation, 
we obtain
\begin{equation}
c_{\boldsymbol k}(t)=A^{(d)}\langle \psi_{\boldsymbol k}
\vert H_{\rm int}\vert \psi_i\rangle\frac{e^{-i(E_D-k-i\gamma/2)t}-1}{E_D-k-i\gamma/2},
\label{sub}
\end{equation}
for $t\geq\Delta t$, where we omit a contribution from the rapid-rise time region $[0,\Delta t]$. 
Substituting Eqs.~(\ref{ansatz}) and (\ref{sub}) into Eq.~(\ref{eqa}), we obtain 
\begin{equation}
-\frac{\gamma}{2}A^{(d)}e^{-\gamma t/2}=-iA^{(d)}\int\frac{d^3k}{(2\pi)^3}
\vert\langle \psi_{\boldsymbol k}\vert H_{\rm int}\vert \psi_i\rangle\vert^2
\frac{e^{-\gamma t/2}-e^{-i(k-E_D)t}}{E_D-k-i\gamma/2}.
\label{gammaeq}
\end{equation}
For $1/E_D\ll t \ll 1/\gamma$, the above equation approximately leads to
\begin{equation}
\gamma=\int\frac{d^3k}{(2\pi)^3}
\vert\langle \psi_{\boldsymbol k}\vert H_{\rm int}\vert \psi_i\rangle\vert^2(2\pi)\delta(k-E_D). 
\end{equation}
The right-hand side results in the same form as Eq.~(\ref{gammarad}) 
and we obtain $\gamma=\Gamma_{\rm rad}$. 
Note that the finite-size correction $A^{(d)}$ is factored out in Eq.~(\ref{gammaeq}) and $\gamma$ is independent of $A^{(d)}$.  
Thus, the natural line width $\gamma$ of the resonance energy is not affected by the finite-size effect. 

The finite-size correction $A^{(d)}$ is determined by the normalization condition
\begin{equation}
\vert c_i(t)\vert^2+\int\frac{d^3k}{(2\pi)^3}\vert c_{\boldsymbol k}(t)\vert^2=1.
\end{equation}
Substituting Eqs.~(\ref{ansatz}) and (\ref{sub}) into this equation, we obtain
\begin{equation}
\vert A^{(d)}\vert^2e^{-\Gamma_{\rm rad} t}
\left(1+\int\frac{d^3k}{(2\pi)^3}\vert\langle \psi_{\boldsymbol k}
\vert H_{\rm int}\vert \psi_i\rangle\vert^2
\frac{\vert 1-e^{-i(k-E_D+i\Gamma_{\rm rad}/2)t}\vert^2}{(E_D-k)^2+\Gamma_{\rm rad}^2/4}
\right)
=1.
\end{equation}
For $1/E_D\ll t \ll 1/\Gamma_{\rm rad}$, the second term in the parentheses becomes $P_{\rm rad}^{(d)}$ approximately. 
Thus, $\vert A^{(d)}\vert^2$ is given by $\vert A^{(d)}\vert^2=1/(1+P^{(d)})$. 
The survival probability $P_i(T)$ and the decay probability $P_{\rm rad}(T)$ for $1/E_D\ll t \ll 1/\Gamma_{\rm rad}$ 
are given by \cite{ishikawa3}
\begin{eqnarray}
P_i(T)&=&\frac{1}{1+P_{\rm rad}^{(d)}}e^{-\Gamma_{\rm rad} T},\\
P_{\rm rad}(T)&=&\frac{P_{\rm rad}^{(d)}}{1+P_{\rm rad}^{(d)}}+\frac{1}{1+P_{\rm rad}^{(d)}}(1-e^{-\Gamma_{\rm rad} T}). \nonumber
\end{eqnarray}
Therefore, the finite-size correction $P_{\rm rad}^{(d)}$ is renormalized as $\tilde{P}_{\rm rad}^{(d)}=P_{\rm rad}^{(d)}/(1+P_{\rm rad}^{(d)})$ in the higher 
order corrections. 
The lifetime of the initial state $\vert i\rangle$ is calculated by
\begin{equation}
T_D=\int_0^\infty P_i(t) dt=\frac{1}{(1+P_{\rm rad}^{(d)})\Gamma_{\rm rad}}.
\end{equation}
For small $P_{\rm rad}^{(d)}$, we obtain $T_D\approx 1/\Gamma_{\rm rad}-P_{\rm rad}^{(d)}/\Gamma_{\rm rad}$. 
Then, the lifetime decreases from $1/\Gamma_{\rm rad}$ by $P_{\rm rad}^{(d)}/\Gamma_{\rm rad}$ because of the finite-size effect. 

\section{Transition amplitude in the second-order perturbation}
\label{transitionam}

In the second-order time-dependent perturbation, the transition amplitude is given by
\begin{equation}
c_f^{(2)}(T)=(-i)^2\int_0^T dt_1\int_0^{t_1} dt_2\langle \psi_f\vert H^I_{\rm int}(t_1)H^I_{\rm int}(t_2)\vert \psi_i\rangle,
\end{equation}
where $H^I_{\rm int}(t)$ is written as
\begin{equation}
H^I_{\rm int}(t)=e^{iH_0 t}\{g\phi(\boldsymbol{r}_A)+g\phi(\boldsymbol{r}_D)\}e^{-iH_0 t}.
\end{equation}
This transition amplitude corresponds to the following process. 
At $t=0$, the initial state is given by $\vert \psi_i\rangle$, a scalar particle is emitted at $t=t_2$, the scalar particle is 
absorbed at $t=t_1$, and the final state $\vert \psi_f\rangle$ is observed at $t=T$. 
Using the initial state $\vert \psi_i\rangle$ and final state $\vert \psi_f\rangle$ given in Sec.~\ref{massless}, 
the transition amplitude is written as
\begin{equation}
c_f^{(2)}(T)=S_A S_D\int d^3r_A d^3r_D\langle A1\vert\boldsymbol{r}_A\rangle\langle\boldsymbol{r}_A\vert A0\rangle
V_{AD}(\boldsymbol{r}_A-\boldsymbol{r}_D)
\langle D0\vert\boldsymbol{r}_D\rangle\langle\boldsymbol{r}_D \vert D1\rangle,
\label{xint}
\end{equation}
where
\begin{equation}
V_{AD}(\boldsymbol{r})=-g^2\int_0^T dt_1\int_0^{t_1} dt_2\Delta(\boldsymbol{r},t_1-t_2)(e^{iE_At_1-iE_Dt_2}
+e^{-iE_Dt_1+iE_At_2}),
\end{equation}
and
\begin{equation}
\Delta(r,t)=\int\frac{d^3 k}{(2\pi)^3}\frac{1}{2k}e^{i\boldsymbol{k\cdot r}-iE_k t-\epsilon k}.
\end{equation}
The terms $S_A$ and $S_D$ are  Franck--Condon factors defined under Eq.~(\ref{initialfinal}). 
$\Delta(r,t)$ is the propagator of the massless scalar field. 
The regularization factor $e^{-\epsilon k}$ is introduced to make the propagator well defined. 
The limit $\epsilon\rightarrow+0$ should be taken after the coordinates integration in Eq.~(\ref{xint}). 
In fact, we obtain the same result by integrating the coordinates before the $k$-integration without the regularization factor. 
Owing to Gaussian form of electron wave functions, the transition amplitude can be analytically calculated as
\begin{equation}
c_f^{(2)}(T)=-ie^{i(E_A-E_D)T/2}S_A S_D\mu_A^i\mu_D^j\frac{\partial^2}{\partial R_i\partial R_j}
\left\{\frac{1}{4\pi R}f(R,T)\right\},
\end{equation}
where $\boldsymbol{R}=\bar{\boldsymbol{r}}_A-\bar{\boldsymbol{r}}_D$ and $f(R,T)=f_1(R,T)+if_2(R,T)$, which is given by
\begin{eqnarray}
f_1(R,T)&=&\frac{2\sigma}{\sqrt{\pi}}\int_0^T dt \frac{\sin(\frac{E_A-E_D}{2}(T-t))}{E_A-E_D}\cos(\frac{E_A+E_D}{2}t)
\{e^{-\sigma^2(t-R)^2}-e^{-\sigma^2(t+R)^2}\},\label{f12}\\
f_2(R,T)&=&-\frac{4\sigma}{\pi}\int_0^T dt \frac{\sin(\frac{E_A-E_D}{2}(T-t))}{E_A-E_D}\cos(\frac{E_A+E_D}{2}t)
\{D(\sigma(t-R))-D(\sigma(t+R))\}.\nonumber
\end{eqnarray}
To obtain the above result, we transformed the variables of integration as $t=t_1-t_2$, $t'=t_1$ and integration 
with respect to $t'$ was carried out.  
The function $D(x)$ is the Dawson function defined by
\begin{equation}
D(x)=e^{-x^2}\int_0^x e^{z^2}dz,
\end{equation}
and $\sigma$ is defined by
\begin{equation}
\sigma^{-2}=\sigma_A^{-2}+\sigma_D^{-2}.
\end{equation}
Using the derivative of the function $f(R,T)$, the transition amplitude is written as
\begin{eqnarray}
c_f^{(2)}(T)&=&iS_A S_D\mu_A^i\mu_D^j\frac{e^{i(E_A-E_D)T/2}}{4\pi R^3} f_{ij}(R,T),\\
f_{ij}(R,T)&=&\{f(R,T)-Rf_R(R,T)\}(\delta_{ij}-3\frac{R_iR_j}{R^2})
-R^2f_{RR}(R,T)\frac{R_iR_j}{R^2},\nonumber
\end{eqnarray}
where $f_R$ and $f_{RR}$ are the first and second partial derivatives of $f(R,T)$ with respect to $R$. 
By using integration by parts, $R$ derivative of $f(R,T)$ generates a factor $E_A-E_D$ or $E_A+E_D$ in 
Eq.~(\ref{f12}). 
These factors result in finite-size corrections that appear mainly in $f_{RR}(R,T)$. 

The function $f(R,T)$ is well approximated for $T>R$ by
\begin{equation}
f_{\rm approx}(R,T)=\frac{2\sigma}{\sqrt{\pi}}\int_0^T dt \frac{\sin(\frac{E_A-E_D}{2}(T-t))}{E_A-E_D}e^{i\frac{E_A+E_D}{2}t}
\{e^{-\sigma^2(t-R)^2}-e^{-\sigma^2(t+R)^2}\}.
\end{equation}
In the following numerical calculations, we use this approximated function and check its validity by 
comparing with the analytic function numerically. 
Note that the integral variable, $t=t_1-t_2$, is the flight time of the scalar particle. 
The Gaussian function in the curly brackets means that the classical trajectory of the massless particle, $t=R$, is dominant 
in the transition amplitude. 
This implies that the scalar particle propagates as a wave packet along the classical trajectory 
and interacts with electrons in the region of size $1/\sigma$. 
In this region, the wave nature of quanta remains and the interaction cannot be ignored. 
Then, finite-size corrections could be generated in this region.

\section{Relation to the standard formula}
\label{relation}

The standard formula for EET is derived by using Fermi's golden rule and the dipole approximation of the transition amplitude. 
To compare the present result to the standard formula, we take the limit $\sigma\rightarrow\infty$ (dipole approximation) and 
$T\rightarrow\infty$ (see Appendix \ref{transitionamplitude} for the limit $\sigma\rightarrow\infty$ with a finite $T$). 
In this limit, the transition probability rate becomes
\begin{equation}
\frac{\vert c_f^{(2)}(T)\vert^2}{T}\rightarrow
\left\vert
S_A S_D\frac{\mu_A^i\mu_D^j}{4\pi R^3}\left\{(1-iE_D R)(\delta_{ij}-3\frac{R_i R_j}{R^2})+(E_D R)^2\frac{R_i R_j}{R^2}\right\}
\right\vert^2
2\pi\delta(E_A-E_D).
\end{equation}
A similar result was obtained in photon interaction theory \cite{MQED,andrews}. 
In photon interaction theory, the last term in the curly brackets becomes a transverse component. 
In the short-range region, the first term in the curly brackets is dominant and reads
\begin{equation}
\left\vert
S_A S_D\frac{\mu_A^i\mu_D^j}{4\pi R^3}(\delta_{ij}-3\frac{R_i R_j}{R^2})
\right\vert^2
2\pi\delta(E_A-E_D).
\label{formula}
\end{equation}
This $R^{-6}$ behavior and the dipole moment dependence coincide with the standard formula of the F\"orster mechanism \cite{fret} 
in which EET occurs through the direct Coulomb term (cf. the Dexter mechanism \cite{dex} due to the exchange Coulomb term).  
This is reasonable because donor and acceptor electrons are well separated in the present model. 
Dirac's delta function in Eq.~(\ref{formula}) represents the resonance between two transition energies of the molecules.  

As seen in the above derivation, the $T$-linear term in the transition probability is taken into account but the $T$-independent
term is ignored in the standard formula. 
Namely, the finite-size correction $P^{(d)}$, which is named after ``diffraction", 
appears in the transition probability $P(T)$ as \cite{ishikawa,ishikawa1,ishikawa2} 
\begin{equation}
P(T)=\Gamma T+P^{(d)},
\label{pd}
\end{equation}
where the transition rate $\Gamma$ and the finite-size correction $P^{(d)}$ are approximately independent of $T$ 
for $T>R$. 
The correction is generated in the region where a scalar particle and electrons are 
overlapping and the interaction cannot be ignored. 
Thus, the finite-size correction $P^{(d)}$ depends on the shape of the wave function and 
appears outside the resonance energy region. 

The transfer time is given by $1/\Gamma$ in the standard theory. 
We define the total transfer time by the time when $P(T)=\Gamma T+P^{(d)}=1$. 
Then, the total transfer time is given by $1/\Gamma-P^{(d)}/\Gamma$. 
We call $T_*=P^{(d)}/\Gamma$ a typical time of the finite-size correction. 
The usual transfer time is defined by $1/\Gamma$, which does not include 
the finite-size correction. 
Note that the typical time is the same form as the reduction time of the lifetime obtained in Sec.~\ref{lifetime}. 
Thus, the total transfer time is shortened by the typical time $T_*$ compared to the usual one. 
At  $T=T_*$, the contribution of the resonance energy region equals the finite-size correction 
by definition. 
At $T<T_*$, the finite-size correction is larger than the contribution of the resonance energy region. 

\section{Numerical calculation for finite-size corrections}
\label{numerical}

In this section, we study finite-size corrections by calculating the function $f_{ij}(R,T)$ numerically. 
Let us decompose $f_{ij}(R,T)$ into the transverse component $f_\perp(R,T)$ and the longitudinal component $f_\parallel(R,T)$ as
\begin{eqnarray}
f_{ij}(R,T)&=&f_\perp(R,T)(\delta_{ij}-\frac{R_i R_j}{R^2})+f_\parallel(R,T)\frac{R_i R_j}{R^2},
\label{component}\\
f_\perp(R,T)&=&f(R,T)-R f_R(R,T),\nonumber\\
f_\parallel(R,T)&=&-2\{f(R,T)-Rf_R(R,T)\}-R^2 f_{RR}(R,T).\nonumber
\end{eqnarray}
It will be shown below that finite-size corrections appear only in the longitudinal component $f_\parallel(R,T)$. 
This is because $f_{RR}(R,T)$ in the longitudinal component includes finite-size corrections mainly. 

Strictly speaking, $\sigma_A$ and $\sigma_D$ depend on the energies $E_A$ and $E_D$. 
However, we treat these quantities as independent parameters for simplicity. 
The energy dependences of $\sigma_A$ and $\sigma_D$ are given in Appendix \ref{energydependence}. 
We use $E_D$ to make physical quantities dimensionless in the following numerical calculation. 

\subsection{$\sigma$-dependence of finite-size corrections}
\begin{figure}[htb]
\begin{center}
\includegraphics[width=9cm]{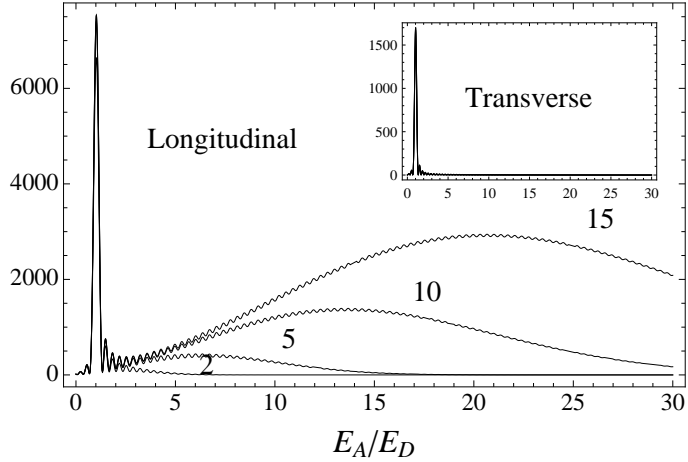}
\caption{Longitudinal component $\vert f_\parallel\vert^2 E_D^2$ for $T E_D=20$, $R E_D=2$, 
$\sigma/E_D=2$, 5, 10 15, and $0<E_A/E_D<30$. 
Inset shows transverse component $\vert f_\perp\vert^2 E_D^2$ for the same parameters. }
\label{fig_bump}
\end{center}
\end{figure}

In Fig.~\ref{fig_bump}, the longitudinal component $\vert f_\parallel(R,T)\vert^2$ and the transverse component $\vert f_\perp(R,T)\vert^2$ 
are plotted for $T E_D=20$, $R E_D=2$, 
$\sigma/E_D=2$, 5, 10, 15, and $0<E_A/E_D<30$. 
As seen in fig.~\ref{fig_bump}, in addition to the resonance peak at $E_A/E_D=1$, a bump appears outside the resonance 
energy region for the longitudinal component. 
On the other hand, the bump is not seen for the transverse component. 
The resonance peak corresponds to the standard formula and 
the bump leads to the finite-size correction $P^{(d)}$. 
The width of the bump is roughly estimated by $\Delta E\approx\sigma$. 

Naively, the present result seems to break energy conservation law, which means that 
the energy expectation value is constant in time. 
The energy expectation value of the initial state at $t=0$ is $E_A^0+E_D^1$; however, finite-size corrections 
due to the bump seem to increase the energy expectation value  of the final state at $t=T$. 
As explained in Sec.~\ref{basic} and \ref{nonstat}, this paradox is resolved by the fact that 
the conserving energy is the expectation value of the total Hamiltonian 
$H_0 + H_{\rm int}$. 
The off-diagonal matrix element of $H_{\rm int}$ between final states (including scalar-particle emission states) 
cancels the energy increase due to the bump. 

The larger $\sigma$ is, the wider and higher the bump is. 
The resonance peak, on the other hand, hardly depends on $\sigma$. 
This means that the resonance peak reflects the particle nature and the bump reflects the wave nature of 
electrons and the scalar particle. 
The bump appears at a higher energy than the resonance energy. 
As shown in Sec.~\ref{spon}, similar behavior appears in the particle emission phenomena \cite{ishikawa,ishikawa1,ishikawa2}. 
This is because the emitted scalar particle can deliver a larger energy than the resonance energy 
for large $\sigma$ (a wave function size in the momentum space). 
In the EET of the present model, the scalar particle is emitted from the donor and delivers a larger energy than the 
resonance energy to the acceptor. 

\subsection{$R$-dependence of finite-size corrections}

\begin{figure}[htb]
\begin{center}
\includegraphics[width=9cm,]{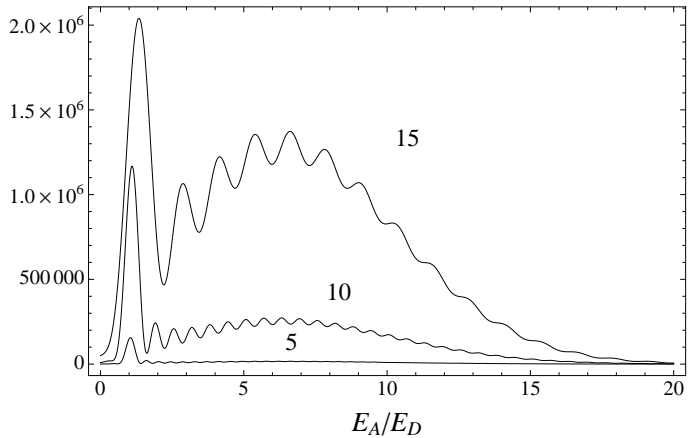}
\caption{The figure shows $\vert f_\parallel\vert^2 E_D^2$ for $TE_D=20$, $RE_D=5$, 10, 15, $\sigma/E_D=5$, and $0<E_A/E_D<20$.}
\label{fig_bump2}
\end{center}
\end{figure}

Let us examine the $R$-dependence of $\vert f_\parallel(R,T) \vert^2$. 
In Fig.~\ref{fig_bump2}, the longitudinal component $\vert f_\parallel(R,T)\vert^2$ is plotted for $T E_D=20$, $R E_D=5$, 10, 15,  
$\sigma/E_D=5$, and $0<E_A/E_D<20$. 
As seen in Fig.~\ref{fig_bump2}, the relative magnitude of the bump to the resonance peak becomes large 
for large $R$. 
This is because finite-size corrections appear mainly in $f_{RR}(R,T)$, which is the long-range term 
of $f_\parallel(R,T)$ in Eq.~(\ref{component}). 
Thus, the resonance peak is dominant in the short-range region and the bump is 
dominant in the long-range region. 

As seen in Fig.~\ref{fig_bump2}, the function $\vert f_\parallel(R,T)\vert^2$ increases for large $R$. 
Note that the transition probability has an overall factor proportional to $R^{-6}$. 
Then, the transition probability decreases rapidly for large $R$. 

\subsection{$T$-dependence of finite-size corrections}
\begin{figure}[htb]
\begin{center}
\includegraphics[width=9cm]{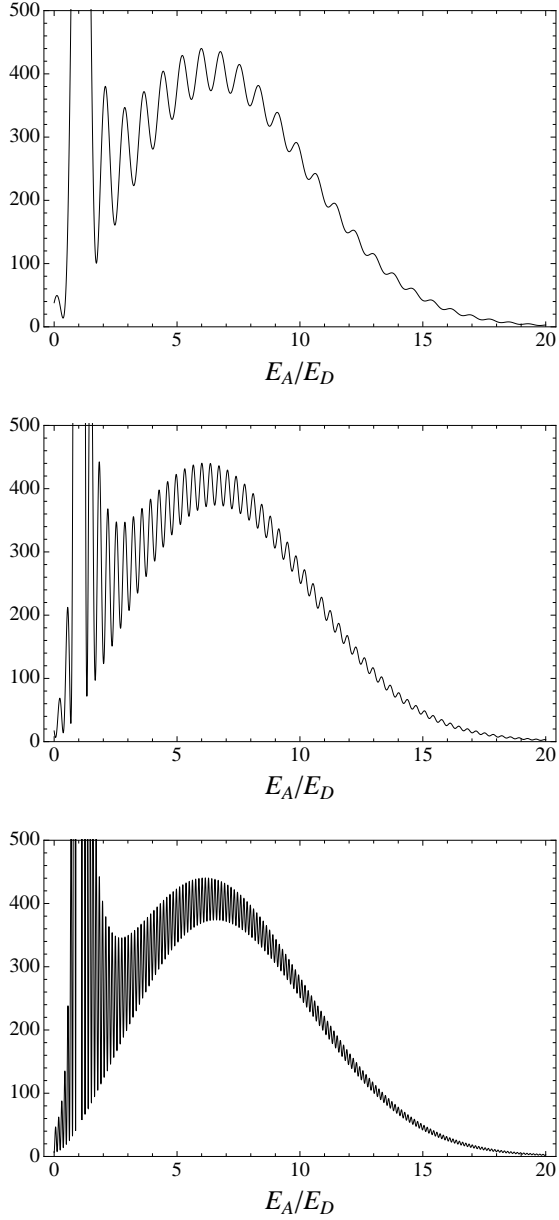}
\caption{The figure shows $\vert f_\parallel\vert^2 E_D^2$ for $T E_D=$10 (top), 20 (middle), 50 (bottom), $R E_D=2$, $\sigma/E_D=5$, and $0<E_A/E_D<20$.}
\label{fig_bump3}
\end{center}
\end{figure}

In Fig.~\ref{fig_bump3}, the longitudinal component $\vert f_\parallel(R,T)\vert^2$ is plotted for $T E_D=10$, 20, 50, $R E_D=2$, 
$\sigma/E_D=5$, and $0<E_A/E_D<20$. 
As seen in Fig.~\ref{fig_bump3}, the bump is hardly dependent on $T$ except for a small rapid oscillation. 
On the other hand, the peak height is proportional to $T^2$ and its width is proportional to $2\pi/T$ (the peaks are not shown in 
Fig.~\ref{fig_bump3}). 
Thus, the resonance peak area is proportional to $T$. 
The characteristics of the resonance peak and the bump are summarized in Table~I. 

\begin{table}[htb]
\caption{Characteristics of the resonance peak and the bump in $\vert f_\parallel\vert^2$. 
The resonance peak and the bump correspond to Fermi's golden rule and the finite-size correction respectively.}
\begin{center}
\begin{tabular}{lccccc}
\hline
&Energy-width&$\sigma$-dependence&$R$-range& Small $T$-behavior&Large $T$-behavior\\
\hline
Resonance peak&Narrow&Small&Short&Small&Linear\\
Bump&Broad&Large&Long&Rapid rise&Constant\\
\hline
\end{tabular}
\end{center}
\end{table}

The transition probability $P(T)$ is given by summing the final states in $\vert c^{(2)}(T)\vert^2$. 
We assume that the state $\vert D0\rangle\vert{\rm nucl}\rangle_{D0}$ is fixed with a transition energy $E_D$ 
and the state $\vert A1\rangle\vert{\rm nucl}\rangle_{A1}$ is distributed with the density of states $D_A(E_A)$, which 
includes the higher excited energy states of the acceptor. 
In the standard F\"orster mechanism, the resonance region of $E_A\approx E_D$ is the main contributor to the EET. 
In this case, it is enough to consider the absorption energy range of the acceptor near the emission energy range of the donor. 
In our mechanism, on the other hand,  the EET of the correlated scalar particle and molecules system has a large contribution from the energy region 
outside the resonance region, $E_A>E_D$. 
Therefore, we must include a wide absorption energy range of the acceptor to calculate the finite-size correction. 
Then, the transition probability is given by
\begin{equation}
P(T)=\int_0^\infty\vert c_f^{(2)}(T) \vert^2 D_A(E_A)dE_A.
\end{equation}
Since the resonance peak and the bump contribute to the energy integral, 
the transition probability behaves as $P(T)=\Gamma T+P^{(d)}$. 
If we ignore the energy dependence of $\sigma$, the Franck--Condon factor, and the density of states, then 
the typical time $T_*=P^{(d)}/\Gamma$ is estimated by calculating the integral
\begin{equation}
\int_0^{E_{\rm max}}\vert f_\parallel(R,T) \vert^2 E_DdE_A= aE_D\, T+b,
\end{equation}
for large $T$, where the cutoff energy $E_{\rm max}$ is introduced. 
The cutoff $E_{\rm max}$ is the upper limit of $E_A$ for which the wave function in Eq.~(\ref{wf}) can be  
regarded as a good approximation for the acceptor electron. 
Dimensionless coefficients $a$ and $b$ are determined numerically. 
Then, the typical time can be estimated by $T_*=b/aE_D$. 
At $T=T_*$, the resonance peak contribution equals the bump contribution. 
Since the bump becomes large for large $\sigma$, the typical time $T_*$ increases with $\sigma$. 

\subsection{Estimation of $T_*$ in LH2 of purple bacteria}\label{estimate}

We use  $RE_D=1$ and $\sigma/E_D=15$. 
For the transition energy $E_D=2\pi/\lambda=2\pi/800$ nm, which corresponds to B800 BChls in LH2 of purple bacteria, 
physical parameters are given by $R=1/E_D=127$ nm, $E_D=1.55$ eV, and $1/\sigma=1/15E_D=8.49$ nm. 
The typical time is estimated as $T_*=$68.9, 6.58, and 1.75 fs for $E_{\rm max}=\infty$, $10E_D$, and $5E_D$, respectively. 
Physical scales for purple bacteria are a few nanometers for the molecule sizes and a few picoseconds 
or a few hundred femtoseconds for the transfer time \cite{p0,p1,p2,p3,p4,p5,p6,p7}. 
Thus, the time-independent finite-size corrections in the transition probability could contribute to a reduction of the total transfer time 
by the estimated typical time.  
For example, if the transfer time, $1/\Gamma$, is 0.2 ps, then $P^{(d)}=\Gamma T_*$ is less than 0.34. 
This means that the back transfer from the acceptor to the donor can be ignored and the lowest-order perturbation is 
a good approximation.

At small $T$, the transition probability rapidly increases to $P^{(d)}$ from 0. 
The width of the bump is roughly estimated as $\Delta E\approx\sigma$. 
Then, the rapid-rise time region of the probability at a small $T$ is estimated by $\Delta t\approx1/\Delta E\approx1/\sigma$. 
Unlike spontaneous emission, the rapid rise of the EET occurs after the flight time $R$ of the scalar particle between the molecules. 
Note that $\Delta t$ corresponds to the overlapping time of electrons and scalar-particle wave functions. 
For $\sigma/E_D=15$, $\Delta t\approx3\times10^{-2}$fs. 
In the transient absorption measurement with 0.1ps light pulses, the rapid-rise behavior must be 
seen as an instantaneous rise from 0 to $P^{(d)}$. 
It is interesting that the similar rapid rise at small $T$ was reported in an EET experiment for the photosynthesis system \cite{p4}. 

Since the bump is broadly distributed, we have to consider the energy dependence of the density of states and 
other physical quantities for a more quantitative analysis. 
The resonance peak contribution mainly comes from the donor and acceptor states with $E_A=E_D$. 
Thus, the standard formula for EET is proportional to the spectral overlap of the donor and acceptor \cite{fret}. 
The bump contribution, on the other hand, comes from the donor and acceptor states with $E_A=E_D+$``bump energy". 
Therefore, we suppose that the finite-size correction due to the bump depends on physical parameters 
(density of states, temperature, etc.), differently from the standard formula. 
For example, spectral peaks of B800 and B850 in LH2 have a gap and the spectral overlap is small. 
Thus, the estimation of the standard F\"orster mechanism for EET from B800 to B850 is small compared to the experiments \cite{p0,p1,p2,p3,p4,p5,p6,p7}. 
Since the contribution of the bump comes from the wide range of each spectrum, 
the finite-size effect of the bump has significance in EET phenomena. 
Furthermore, the finite-size correction could be a finite value even if the spectral overlap of the donor and acceptor is zero. 
In other words, EET could occur between two molecules with very different excitation energies due to the bump. 
Investigation of this finite-size effect in general molecular systems is our important future work. 
 
\section{Summary and Discussion}
\label{summary}

In the standard theory of EET, the F\"orster mechanism is used for resonance energy transfer in the weak-coupling regime. 
Many aspects of EET in the photosynthesis system can be understood through the F\"orster mechanism. 
However, we have not yet obtained a full understanding of the extremely high efficiency of EET in some systems, such as the light-harvesting complex of purple bacteria. 
Experimental and theoretical research has revealed that interplay between exciton modes and the environment plays an important role 
in the high efficiency of EET in the photosynthesis system \cite{p0,p1,p2,p3,p4,p5,p6,p7,ishizaki,schulten}. 
In this paper, we proposed a new mechanism that enhances EET by the finite-size effect outside the resonance energy region. 

We investigated a simple model in which the energy is transferred by a massless scalar particle as an analog of a photon.  
It is found that the interaction energy due to the overlapping quantum waves causes the large finite-size effect. 
The transition amplitude with a finite $T$ (time interval) and $1/\sigma$ (wave-function size) exhibits the standard resonance peak 
and the broad bump outside the resonance energy region. 
This bump leads to the finite-size correction $P^{(d)}$. 
The bump has a width proportional to $\sigma$ and is hardly dependent on $T$ for a large $T$. 
At a small $T$, the correction term rapidly increases to $P^{(d)}$ from 0. 
This rapid rise time-region is very short and roughly given by $\Delta t\approx 1/\sigma$, 
which corresponds to the overlapping time of electrons and scalar-particle wave functions. 

We used time-dependent perturbation theory and found that the finite-size effect appears at the tree level. 
The continuum states of the scalar particle emitted from the donor lead to the broad bump and 
rapid rise in the transition probability.  
Since the rapid-rise time region is very short, back transfer can be ignored in the perturbation theory,  
which is usually used in the weak coupling regime. 
We verified the validity of the perturbation theory by calculating the higher-order effect. 
It was shown that the natural line width due to the lifetime effect is independent of the finite-size effect. 

We estimated a typical time for the finite-size correction using physical parameters of purple bacteria. 
The estimated value was compared to the transfer time of purple bacteria. 
Then, we conclude that the total transfer time could be reduced 
by the time-independent correction term in the transition probability in the photosynthesis system. 
Usually, the transfer time is estimated by the inverse of the transfer rate. 
However, the total transfer time introduced in this paper cannot be estimated by the transfer rate because of 
the constant correction term in the transition probability. 
We expect a novel experiment to observe the constant correction term directly in the near future. 

Naturally, to make a more quantitative comparison with a realistic system, we have to 
study EET by using a photon instead of a scalar particle. 
We predict that finite-size corrections to Fermi's golden rule are generated by the interaction energy due to 
overlapping between the photon and electron waves. 
We hope that the finite-size effect due to the wave nature of quanta sheds a new light on the photosynthesis and related 
molecular phenomena.

\begin{acknowledgments}
The present work was partially supported by a Grant-in-Aid for Scientific Research (Grant No.~24340043), 
a Grant-in-Aid for Scientific Research (C) 15K00524, and JSPS KAKENHI Grant No. JP15H05885 (J-Physics). 
T. Yabuki thanks Prof. Shigenori Tanaka and Prof. Kuniyoshi Ebina 
for their useful comments and encouragement throughout this work. 
\end{acknowledgments}

\appendix
\section{Transition amplitude in the dipole approximation}
\label{transitionamplitude}
In this appendix, we calculate the limit of $\sigma\rightarrow\infty$ (dipole approximation). 
The function $f(R,T)=f_1(R,T)+if_2(R,T)$ for $\sigma\rightarrow\infty$ with a finite $T$ is given by
\begin{eqnarray}
f_1(R,T)&\rightarrow&\frac{2\sin(\frac{E_A-E_D}{2}(T-R))}{E_A-E_D}\cos(\frac{E_A+E_D}{2}R)\theta(T-R),\\
f_2(R,T)&\rightarrow&-\frac{2}{\pi}\int_0^T dt\left({\cal P}\frac{1}{t-R}-\frac{1}{t+R}\right)
\frac{\sin(\frac{E_A-E_D}{2}(T-t))}{E_A-E_D}\cos(\frac{E_A+E_D}{2}t).\nonumber
\end{eqnarray}
where $\cal P$ stands for the principal value. 
For large $T$ limit, the function $f(R,T)$ is given by
\begin{equation}
f(R,T)\rightarrow \frac{2\sin(\frac{E_A-E_D}{2}T)}{E_A-E_D}e^{i\frac{E_A+E_D}{2}R}.
\end{equation}
In the standard theory, this is approximated by using Dirac's delta function as
\begin{equation}
\frac{2\sin(\frac{E_A-E_D}{2}T)}{E_A-E_D}e^{i\frac{E_A+E_D}{2}R}=2\pi\delta(E_A-E_D)e^{iE_D R},
\end{equation}
for the large $T$ limit. 
Note that the validity of usage of Dirac's delta function must be verified carefully in applying to a physical system. 

\section{Energy dependence of $\sigma_A$ and $\sigma_D$}
\label{energydependence}
In this appendix, we study the energy dependence of $\sigma_A$ and $\sigma_D$. 
We have the commutation relations
\begin{equation}
[\boldsymbol{r}_A,H_0]=i\boldsymbol{p}_A/m,\ [\boldsymbol{r}_D,H_0]=i\boldsymbol{p}_D/m.
\end{equation}
Using these relations and Eqs.~(\ref{eigen}) and (\ref{wf}), we can easily obtain 
\begin{equation}
\sigma_A^2=mE_A,\ \sigma_D^2=mE_D.
\end{equation}
Thus, $\sigma_A$ and $\sigma_D$ increase for large $E_A$ and $E_D$ respectively. 
These cause the dipole moments and $\sigma$ to have an energy dependence. 
Then, the asymptotic behavior of the transition probability is changed by these effects. 
However, we checked that the conclusion obtained in the present paper was not changed.


\begin{thebibliography}{99}
\bibitem{dirac} P. A. M. Dirac, Proc. R. Soc. London, Ser. A {\bf 114}, 243 (1927). 
\bibitem{oppen} J. R. Oppenheimer, Phys. Rev. {\bf 60}, 158 (1941).
\bibitem{oppen2} W. Arnold and J. R. Oppenheimer, J. Gen. Physiol. {\bf 33}, 423 (1950). 
\bibitem{fret} T. F\"orster, Ann. Phys. (Berlin) {\bf 437}, 55 (1948).
\bibitem{dex} D. L. Dexter, J. Chem. Phys. {\bf 21}, 836 (1953). 
\bibitem{p0} X. Hu, T. Ritz, A. Damjanovic, and K. Schulten, J. Phys. Chem. B {\bf 101}, 3854 (1997). 
\bibitem{p1} V. Sundstr\"om, T. Pullerits, and R. van Grondelle, J. Phys. Chem. B {\bf 103}, 2327 (1999). 
\bibitem{p2} K. Mukai, S. Abe, and H. Sumi, J. Phys. Chem. B {\bf 103}, 6096 (1999). 
\bibitem{p3} G. D. Scholes and G. R. Fleming, J. Phys. Chem. B {\bf 104}, 1854 (2000).
\bibitem{p4} J. L. Herek. N. J Fraser, T. Pullerits, P. Martinsson, T. Pol\'ivka, H. Scheer, R. J. Cogdell, and V. Sundstr\"om, 
Biophys. J. {\bf 78}, 2590 (2000). 
\bibitem{p5} A. Kimura and T. Kakitani, J. Phys. Chem. B {\bf 107}, 7932 (2003).
\bibitem{p6} S. Jang, M. D. Newton, and R. J. Silbey, Phys. Rev. Lett. {\bf 92}, 218301 (2004). 
\bibitem{p7} Y. C. Cheng and R. J. Silbey, Phys. Rev. Lett. {\bf 96}, 028103 (2006). 
\bibitem{ishizaki} A. Ishizaki and G. R. Fleming, J. Chem. Phys. {\bf 130}, 234111 (2009). 
\bibitem{schulten} J. Str\"umpfer, M. \c{S}ener, and K. Schulten, J. Phys. Chem. Lett. {\bf 3}, 536 (2012). 
\bibitem{ishikawa} K. Ishikawa and Y. Tobita, Prog. Theor. Exp. Phys. {\bf 2013}, 073B02 (2013).
\bibitem{ishikawa1} K. Ishikawa and Y. Tobita, Ann. Phys. {\bf 344}, 118 (2014). 
\bibitem{ishikawa2} K. Ishikawa, T. Tajima, and Y. Tobita, Prog. Theor. Exp. Phys. {\bf 2015}, 013B02 (2015).
\bibitem{von} J. von Neumann, {\it Mathematical Foundations of Quantum Mechanics} (Princeton University Press, Princeton, 1955).
\bibitem{ww} V. Weisskopf and E. Wigner, Z. Phys. {\bf 63}, 54 (1930).
\bibitem{ww2} V. Weisskopf and E. Wigner, Z. Phys. {\bf 65}, 18 (1930).
\bibitem{ishikawa3} K. Ishikawa, T. Nozaki,  M. Sentoku, and Y. Tobita, [arXiv:1405.0582 [hep-ph]].
\bibitem{ishikawa4} K. Ishikawa and Y. Tobita, [arXiv:1607.08522 [hep-ph]].
\bibitem{MQED} D. P. Craig and T. Thirunamachandran, {\it Molecular Quantum Electrodynamics} (Dover Publications, New York, 1998).
\bibitem{andrews} D. L. Andrews and D. S. Bradshaw, Ann. Phys. (Berlin) {\bf 526}, 173 (2014). 
\end{thebibliography}
\end{document}